# The sunrise amplitude equation applied to an Egyptian temple

**Amelia Carolina Sparavigna**
Institute of Fundamental Physics and Nanotechnology
Department of Applied Science and Technology
Politecnico di Torino, C.so Duca degli Abruzzi 24, Torino, Italy

*An equation, fundamental for solar energy applications, can be used to determine the sunrise amplitude at given latitude. It is therefore suitable for being applied to archaeoastronomical calculations concerning the orientation of towns, worship places and buildings. Here we discuss the cases of the Karnak worship complex, at Luxor, Egypt, and of the Great Temple of Amarna, and their orientation toward the sunrise.*

Recently the equation of the sunrise amplitude had been applied to study the orientation of the ancient Roman core of Torino [1,2]. The sunrise amplitude is the angle given by the direction of the rising sun with respect to the cardinal East-West direction. An equation describes this angle as a function of latitude and declination was used by P.I. Cooper in his paper on solar stills [3], a paper which is a fundamental reference for solar energy applications [4]. For reader's convenience, declination, hour angle and sunrise amplitude are reported in the Appendix.

In [1], it is discussed how sometimes the Romans, and before Greeks and Etruscans, used to orient their towns with the sunrise on the day of the foundation. In fact, Torino has the main street oriented according to the sunrise during winter. After this main direction and its perpendicular had been determined, the rectangular area of the new town was subdivided in a chessboard of "insulae", the modern house-blocks [5]. Therefore, in towns of Roman foundation, buildings and worship places were oriented according to this chessboard.

The orientation of buildings, besides its importance in reducing energy consumption, is discussed in many references for its symbolic role. For instance, Christian churches generally point eastwards. But, as told in Ref.6, it is better to remark that "eastwards" does not necessarily mean "due East". Therefore, the orientations of many churches deviate from the cardinal East-West direction by considerable amounts. The question is then how was determined the direction that the worship place is facing. Various suggestions have been proposed, as reported in [6,7], the main ones are related to the rising sun, in particular, it is guessed that the churches are oriented according to the sunrise on the day when their foundations had been laid. There is historical evidence of this practice for the churches built during the seventeenth and eighteenth century in England [6].

When we have no documents or other information on the foundation of buildings and towns, we can check whether an orientation according to the sunrise is possible or not, using the sunrise amplitude equation. That is, measuring the orientation with respect to the East-West direction and using the aforementioned equation we can determine the day of foundation [1,8].

The sunrise orientation is quite older: we can find it for some Neolith and Bronze Age structures, such as the Goseck circle [9], and in the temples of the ancient Egypt [10,11]. In this paper, let us apply the sunrise amplitude equation to an ancient Egyptian worship place, that of the Great Temple of Karnak. This temple complex is composed by several buildings. The complex creation began in the reign of Senusret I (Sesostris I) in the Middle Kingdom. He ruled from 1971 BC to 1926 BC, The area around Karnak was the main place of worship of the god Amun.

As discussed by D. Furlong [10], the main directional orientation of the vast temple of Amun, which stands on the east bank of the Nile, corresponds with a mid-summer sunset on a level horizon. "This is what Sir Norman Lockyer suggested, in his book The Dawn of Astronomy first published in 1894." [12] Let us check the orientation using the sunrise amplitude given in the Appendix, as a function of the days after the spring equinox, at Luxor latitude and compare it with the angle we can obtain from the Google Maps.

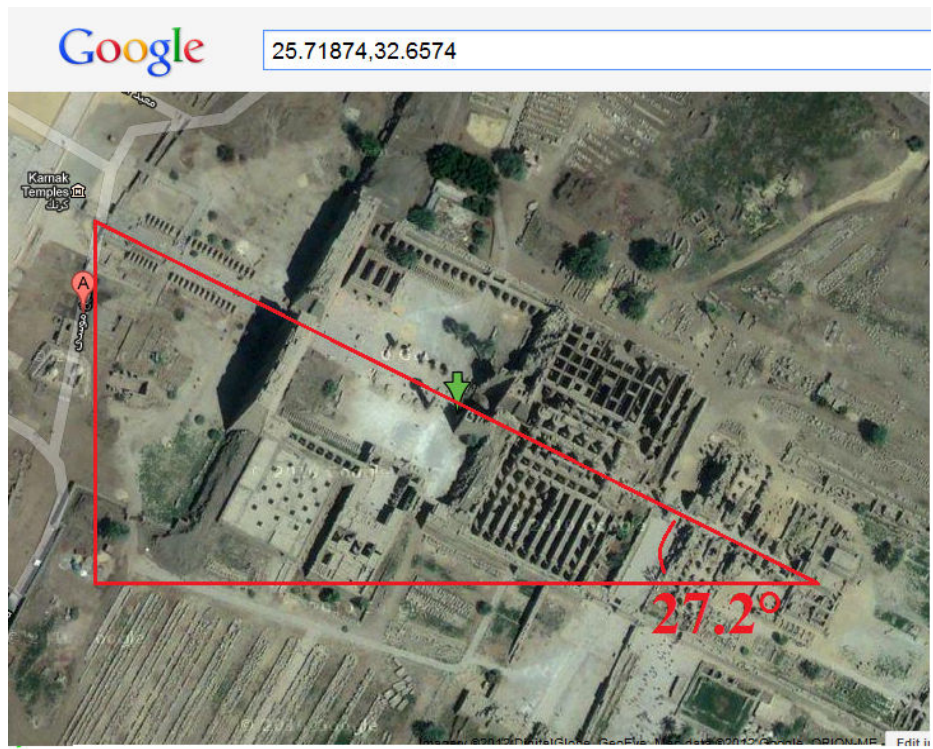

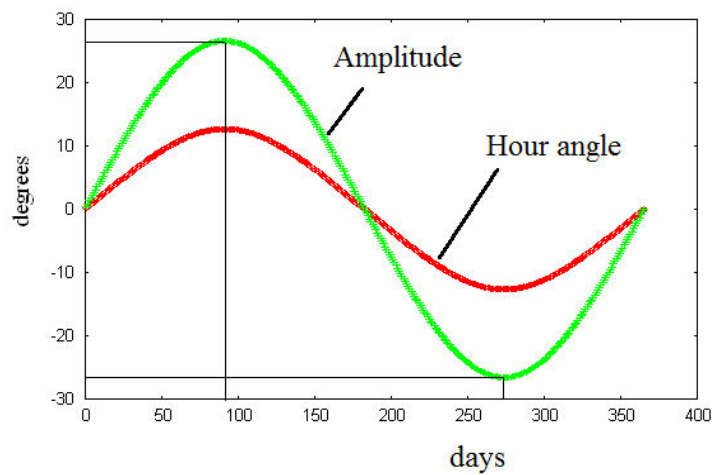

*Fig.1. In the upper part of the figure we see the satellite map of Karnak. Using two sides of the right-angled triangle (one is the East-West direction) we find the angle. In this case it is of 27.2° (negative). The hypotenuse is the axis of the temple. In the lower part, we see the sunrise amplitude (green) as a function of the days after the spring equinox, evaluated at Luxor latitude. We immediately see that the angle corresponds to solstices: the winter solstice in the case of the sunrise and the summer solstice in the case of the sunset.*

Using the satellite maps, we can immediately evaluate the angle the axis of the temple is forming with the cardinal East-West direction (see Fig.1). After plotting the amplitude, we see that the angle corresponds to the sunrise of winter solstice and the sunset of summer solstice. Probably, the ancient Egypt had some rituals for the foundation of temples, and these rituals passed to ancient Greece and Rome.

Let us consider another case, that of the Great temple of Amarna. This temple was built by Akhenaten, a king of the Eighteenth dynasty of Egypt, who ruled for 17 years and probably died between 1336-34 BCE. This king was known before the fifth year of his reign as Amenhotep IV.

Akhenaten abandoned the traditional Egyptian polytheism and introduced a worship centered on the Aten, the Sun, and changed his name accordingly. After his death, the traditional religious practice was restored and his name cancelled from Egyptian history, until the discovery of Amarna, the city founded and built by this king. The excavations at Amarna were lead by Flinders Petrie [13]. On the east bank of the Nile River, the city was built on a virgin site [13], as the new capital of Egypt, dedicated to the worship of Aten.

The earliest dated stelae from Akhenaten's new city is telling that "His Majesty mounted a great chariot of electrum, like the Aten when He rises on the horizon and fills the land with His love, and took a goodly road to Akhetaten, the place of origin, which [the Aten] had created for Himself that he might be happy therein." What is written on the stelae is a reference to the sun rising from the horizon, and therefore, we have to guess that the orientation of Amarna buildings must have a strong relation with the sun motion over horizon. In fact, the name of Amarna is "Akhetaten", the "Horizon of Aton" [10,11].

Recently [14] I have investigated with Google Maps and image processing the Amarna site. Quite visible in the satellite images, after processing, are the Great Temple, the Small Temple and the Desert Altar. The orientations of these worship places are the same. Let us consider the Great Temple. The construction of this temple was accomplished in several steps [13]. First of all, there was a ceremony at the site followed by the building of the "temenos" wall, the enclosure which is surrounding a huge area of 229 m x 730 m. We can see it in Figure 2. The wall is forming an angle with the cardinal East-West direction, and we can easily obtain that this angle is of 13.4 degrees.

Let us compare it with the sunrise amplitude at Amarna latitude. As previously told, the sunrise amplitude is an angle determined by the rising sun with respect to the cardinal E-W direction. As in [2-4], the sunrise amplitude Z (in degrees) is given as a function of the latitude $\varphi$ and the solar declination $\delta$ (see Appendix), which can be represented as a function of the number of days after the spring equinox.

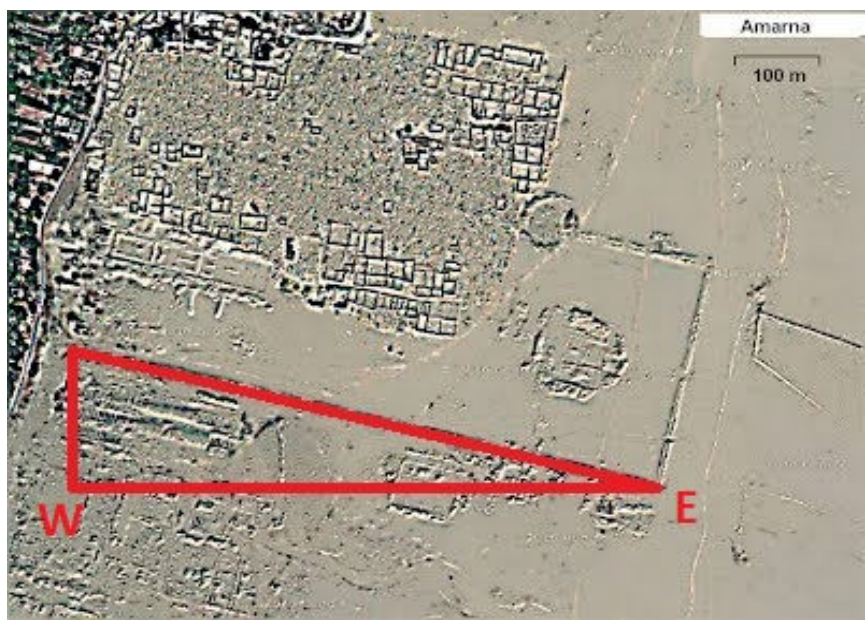

*Fig.2 The angle the wall of the Great Temple enclosure is forming with the cardinal W-E direction is of 13.4 degrees (negative), this value corresponding to the sunrise amplitude of the winter solstice.*

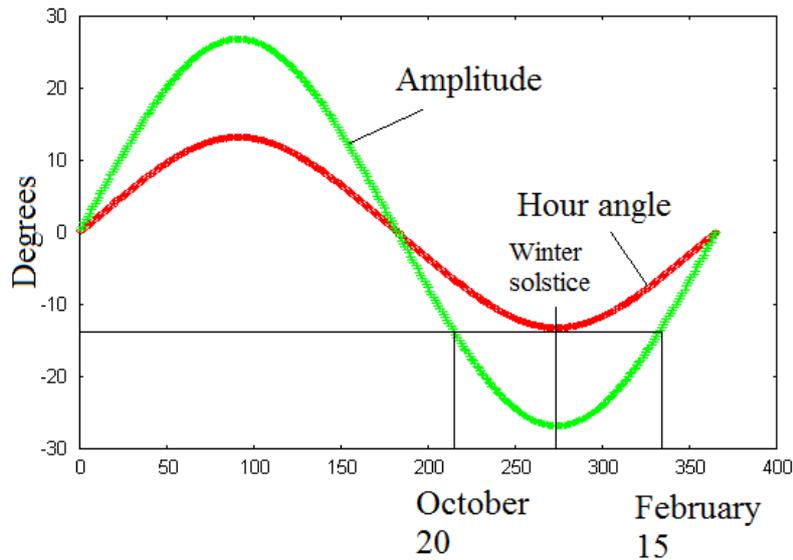

*Fig.3 Sunrise amplitude at Amarna as a function of the days after spring equinox. The angle is in degrees.*

Plotting Z we have Fig.3: in this figure, the hour angle ω is given for comparison too. We see that the angle of the wall enclosure is according to the sunrise amplitude on two days: October 20 and February 15, approximately. Or it could be oriented with the sunset and therefore we have other two days. But, in agreement with what the stelae is telling, that the king oriented his worship place with the rising Sun, the two days oriented according to the sunset are not considered. There is another quite interesting fact concerning the Great Temple. If we observe the map of the region near Amarna (please see the first image of Ref.14), the temple is facing the Royal Wadi, the royal burial place of Amarna.

To decide between the two rising suns, we need more information. As we did in the case of the Roman town of Timgad, we can search for an agreement with dates relevant during the founder's life (in the case of Timgad, emperor Trajan), or for the Egyptian religion or civil calendar [15]. In the case of the king of Akhetaten, the Sun Horizon, I like to imagine that he founded the temple to commemorate the day when he became Akhenaten, in agreement with the words written on the stelae. Let us remark that the sunrise angle Z and satellite imagery confirm the possibility of a foundation according to a sunrise amplitude.

In astronomic terms [16], the coordinate system used in this approach is the horizontal one. It needs the visible horizon free from hills or mountains to be applied during the ritual, by an observer that had to find the sunrise location. In the case that this is impossible, let us consider an alternative approach, based on the use of the equatorial coordinate system. From the plots in Fig.3, we see that it is quite interesting the behavior of the hour angle, which on winter solstice is coincident with the angle of the Great Temple orientation. This means that the orientation of the enclosure, and also of the Deset Altar, could had been decided as a local image of the sky, in a literal sense, where the structures are in agreement with equatorial angles. The maximum value of the hour angle is 13.4 degrees (negative for the winter solstice). At the winter solstice this hour angle turns out to be the angle of the temenos. In this case, Akhenaten decidde to represent the sky on Earth, in a different manner from the priests of Karnak. The same situation, a correspondence of the hour angle with the orientation of a structure, happens for the main street of the Roman Torino, where its orientation coincides with the hour angle at the winter solstice.

As we have seen in the discussed cases, Karnak and Amarna, the sunrise amplitude equation, combined with an image processing of free satellite images, can be quite useful for archaeological and archaeoastronomical studies, allowing everybody to investigate the orientations of local structures before planning any direct on site investigation on the winter solstice. Other sites are under investigation to check the horizontal orientation and to see whether orientation with equatorial coordinates is possible or not.

**Appendix**

We need to know the latitude and the declination. In [2-4] it had been proposed and used a formula for the declination as a function of the days after the spring equinox, as reported in the following table. The declination, with the hour angle ω, are the coordinates in the equatorial coordinate system [16]. To have the sunrise amplitude Z, that is the direction of the sun on the observer's horizon, we use the last equation in the table, as that given in http://www.titulosnauticos.net/astro/

latitude $\varphi$

declination (in radians)

$$\delta = arcsin(0.4 \cdot sin(2\pi n / 365))$$

$n$ = number of days after the spring equinox

hour angle $\omega$

$$\omega = \frac{360°}{2\pi} arccos(-tan\varphi \cdot tan\delta) - 90°$$

sunrise amplitude

$$Z = 90° - \frac{360°}{2\pi} arccos(sin\delta / cos\varphi)$$

A quite useful visual representation of the direction of sunset, noon, and sunrise for a location is provided at the site http://www.sollumis.com/. This is implemented on Google Maps, suitable for any location on the map.